\documentclass[aps,pra,reprint,groupedaddress,showpacs]{revtex4-1}
\usepackage{amsmath}
\usepackage{amssymb}
\usepackage{color}
\usepackage[colorlinks,bookmarks=false,citecolor=red,linkcolor=blue,urlcolor=blue]{hyperref}
\usepackage{graphicx}
\usepackage{subfigure}
\def\ahat{\hat{a}}

\def\chat{\hat{c}}

\def\phihat{\hat{\phi}}
\def\psihat{\hat{\psi}}
\def\chihat{\hat{\chi}}

\def\nhat{\hat{n}}
\def\Dhat{\hat{D}}

\def\Phat{\hat{P}}
\def\Hhat{\hat{H}}
\def\Uhat{\hat{U}}

\def\calL{\mathcal{L}}

\begin{document}
\title{Quantum Ising dynamics and Majorana-like edge modes in the Rabi lattice model}
\author{Brijesh Kumar}
\email{bkumar@mail.jnu.ac.in}
\author{Somenath Jalal}
\affiliation{School of Physical Sciences, Jawaharlal Nehru University, New Delhi 110067, India}
\date{\today}
\begin{abstract}
The atomic dipoles in the Rabi lattice model exhibit quantum Ising dynamics
in the limit of strong atom-photon interaction. It governs the para- to ferro-electric phase transition in the ground state. On an open chain, it implies the existence of two Majorana-like, albeit topologically unprotected, edge modes in the ordered phase. The relation $\rho^x_{1L}=p^8$ between the end-to-end dipole correlation, $\rho^x_{1L}$, and the spontaneous polarization, $p$, is proposed as an observable signature of these edge modes. The DMRG calculations on the one-dimensional Rabi lattice support the strong coupling quantum Ising behavior, and correctly yield the proposed end-to-end dipole correlation. The conditions which protect the edge modes against the adverse perturbations are also identified.
\end{abstract}
\maketitle

{\em Introduction.}
The lattice models of quantum cavities have received much attention recently, for emulating  the Bose-Hubbard model by exhibiting the Mott-insulator to superfluid quantum phase transition for photons~\cite{Tomadin.Fazio.10,Hartmann,Greentree.06,Littlewood,Blatter}. A quantum cavity 
refers to an interacting matter-radiation system inside a high-Q resonator (cavity)~\cite{Haroche.Raimond}. The simplest of it can be modeled as a two-level atom interacting with a single mode of quantized radiation. It is believed that such lattice models may be realized, for example, by engraving an array of cavities in a photonic band-gap material~\cite{Greentree.06,QD.Photonic}, or as a circuit-QED system~\cite{CircuitQED.Girvin,Houck.Koch}, and will be useful in studying quantum many-body problems, like the cold atoms in optical lattices~\cite{ColdAtoms}.

The superfluid to insulator transition in the ground state of 
the Bose-Hubbard (BH) model is an example of a U(1) symmetry-breaking quantum phase transition (QPT), driven by the competition between kinetic energy and local repulsion~\cite{BH_Fisher, BH_Sheshadri}. Another case which beautifully typifies QPTs is the quantum Ising (QI) model~\cite{QIsing.Pfeuty,QIsing.deGennes}. It describes an Ising system in a field transverse to the Ising axis, and exhibits a Z$_2$ symmetry-breaking transition. The QI model has been used to study a variety of physical problems~\cite{QIsing.deGennes,Columbite1,quench.QI}. It has also 
served as an important point of reference in the general understanding of  QPTs~\cite{QPT.Sachdev}.
 
The one-dimensional (1D) spin-1/2 QI model with only nearest-neighbor interaction is exactly soluble using Jordan-Wigner fermionization~\cite{QIsing.Pfeuty}. 
For the 1D QI model in the fermionized form, viz. a p-wave superconducting chain, Kitaev observed that the two ends of an open chain carry a Majorana fermion each~\cite{Kitaev.FreeMajorana}. A fermion which is Hermitian (an anti-particle of itself) is called a Majorana fermion~\cite{Kitaev.FreeMajorana, Wilczek.Majorana, note.Majorana}. There has always been a great interest in finding the Majorana fermions in nature. In recent times, the proposed role of Majorana fermions in topological quantum computation~\cite{Kitaev.FreeMajorana,Nayak.RMP} has invigorated their search in condensed matter systems~\cite{Mourik.Majorana,Alicea,Majorana.Beenakker}. Notably, the Majorana fermions in Kitaev's quantum wire are topologically protected from local perturbations~\cite{Kitaev.FreeMajorana,Topo_QC_Kitaev}. This is not so for the Majorana fermion modes in the QI chain. To keep this distinction, we call the Majorana modes in the QI chain as `Majorana-like edge modes', or simply the `edge modes', as the term `Majorana modes' is often used for the topologically protected ones.

In this paper, we present a case for the realization of Majorana-like edge modes, through QI dynamics, in the Rabi lattice model (RLM). By RLM we mean an array of the Rabi quantum cavities where in each cavity a two-level atom (or a spin-1/2) is coupled via dipole interaction to a single photon mode, and the inter-cavity coupling causes photon hopping. We show that the RLM in the limit of strong atom-photon coupling rigorously tends to the QI model, that governs para- to ferro-electric QPT therein. It also gives two Majorana-like edge modes in the ordered phase of the 1D RLM. We propose $\rho^x_{1L}=p^8$ as a signature of these edge modes, where $p$ is the bulk polarization and $\rho^x_{1L}$ is the end-to-end correlation of atomic dipoles in the ground state on an open chain of $L$ cavities. We identify conditions that save the edge modes from the detrimental perturbations such as a stray electric field. We support our findings by doing DMRG (density matrix renormalization group~\cite{Schollwock2005}) calculations  in 1D. 

{\em Rabi lattice model.}
A Rabi quantum cavity is a minimal atom-photon problem described by the Rabi model: $\Hhat_R=\omega (\ahat^\dag\ahat+\frac{1}{2})+\frac{\varepsilon}{2}\sigma^z +\gamma \sigma^x(\ahat^\dag+\ahat)$, where $\omega$ is the photon energy and $\varepsilon$ is the atomic transition energy. The dipole interaction, $\vec{p}\cdot\vec{E}$, is written as $\gamma\sigma^x (\ahat^\dag+\ahat)$, where $\gamma$ denotes the atom-photon coupling, $\ahat$ ($\ahat^\dag$) is the photon annihilation (creation) operator, and the Pauli operator $\sigma^x = |e\rangle\langle g| + |g\rangle\langle e|$, measures the atomic dipole, $\vec{p}$~\cite{Scully.Zubairy}. Moreover, $\sigma^z=|e\rangle\langle e| - |g\rangle\langle g|$, and $|e\rangle\langle g|=\sigma^+$. The kets $|g\rangle$ and $|e\rangle$ denote the atomic ground and excited levels, respectively. The $\Hhat_{R}$ is deceptively simple, with no closed-form exact solution yet, although it has been studied extensively~\cite{RabiModel.Graham, RabiModel.Reik, RabiModel.Braak}. 
 A popular variant of this by Jaynes and Cummings is however exactly soluble~\cite{Jaynes.Cummings}. 

In the Jaynes-Cummings (JC) model, the atom-photon 
interaction is taken as $\gamma(\sigma^+\ahat + \sigma^-\ahat^\dag)$, by dropping the counter-rotating terms, $\sigma^+\ahat^\dag + \sigma^-\ahat$, from $\gamma\sigma^x(\ahat^\dag+\ahat)$. This so-called rotating-wave approximation (RWA) is applicable close to resonance ($\omega \simeq \varepsilon$) for weak couplings ($\gamma\ll\omega$). 
It leads to the conservation of `polariton' number, $\ahat^\dag\ahat + \sigma^+\sigma^-$, in the JC model, and makes it analytically soluble. This continuous $U(1)$ symmetry is absent in the Rabi model, which only has a discrete parity symmetry, $-\chihat \sigma^z$, where $\chihat=(-1)^{\nhat}$ and $\nhat=\ahat^\dag\ahat$.
 
The studies that find the superfluid-insulator QPT for photons have focussed on the arrays of JC cavities~\cite{Tomadin.Fazio.10}. Since the RWA and photon hopping together conserve the total polariton number, the JC lattice 
models exhibit a spontaneous $U(1)$ symmetry-breaking transition akin to the BH model. However, inside a high finesse cavity with strong atom-photon interaction, the counter-rotating terms would inevitably express. Recent studies indicate a good possibility of going beyond RWA into the strong coupling regime~\cite{Rabi.realize1,Rabi.realize2}. It is therefore important to understand the physics of cavity lattice models without RWA. Hence, we study the RLM,  
written below. 
\begin{equation}
\Hhat = \sum_l \Hhat^{ }_{R,l} -t\sum_{l,\delta} \left( \ahat^\dag_l\ahat^{ }_{l+\delta} + \ahat^\dag_{l+\delta}\ahat^{ }_l\right)
\label{eq:H}
\end{equation}
Here, $\Hhat_{R,l}$ denotes the $l^{th}$ Rabi cavity in the lattice. The second term in $\Hhat$ is the nearest-neighbor photon hopping with amplitude $t$. (The rates of various incoherent losses are assumed to be $\ll \gamma$, $\omega$, $\varepsilon$ and $t$.)
 
In the RLM [that is, $\Hhat$ of Eq.~(\ref{eq:H})], the parity of individual cavities is not conserved, because the photon hopping incessantly changes the photon numbers in the cavities which causes the the local parities to fluctuate between odd and even. However, the global parity, $\prod_l (-\chihat_l\sigma^z_l$), is still conserved. As the global parity is a discrete symmetry, a QPT in the RLM would occur by breaking it, and not be accompanied by gapless  
excitations in the `ordered' phase (also noted in Ref.~\onlinecite{RabiLattice.ZHeng}).  
In fact, we find the Rabi lattice to exhibit quantum Ising transition.
Below we establish this by transforming $\Hhat$ to a form wherein the underlying QI dynamics becomes evident. 

{\em Quantum Ising dynamics.} 
Consider $\sigma^x_l (\ahat^\dag_l +\ahat^{ }_l)$. Let  $\sigma^x_l$ be absorbed into  
$(\ahat^\dag_l+\ahat^{ }_l)$ by the unitary transformation, 
$\Uhat=\prod_l (\Phat^+_l+\Phat^-_l\chihat_l)$, where $\Phat^\pm_l=(1\pm\sigma^x_l)/2$.
Under this transformation, the $\nhat_l$ and $\sigma^x_l$ stay unchanged, while $\Uhat^\dag \ahat_l \Uhat = \sigma^x_l\ahat_l$ and $\Uhat^\dag \sigma^z_l \Uhat = \chihat_l\sigma^z_l$. Thus, in the transformed Rabi cavity, the dipole interaction turns into the displacement field, $\gamma (\ahat^\dag_l+\ahat^{ }_l)$, the parity is given by $\sigma^z_l$, and $\sigma^x_l$ continues to be the atomic dipole.

In the Rabi lattice model transformed under $\Uhat$,
\begin{eqnarray}
&& \Uhat^\dag \Hhat \Uhat = \Hhat_1 \nonumber = \sum_l \left[\omega\left(\nhat_l+\frac{1}{2}\right)+\gamma\left(\ahat^\dag_l+\ahat^{ }_l\right) \right] \nonumber\\
&&+\frac{\varepsilon}{2}\sum_l\chihat_l\sigma^z_l -t\sum_{l,\delta}\sigma^x_{l}\sigma^x_{l+\delta} \left( \ahat^\dag_l\ahat^{ }_{l+\delta} + \ahat^\dag_{l+\delta}\ahat^{ }_l \right), \label{eq:H1}
\end{eqnarray}
the atom-photon interaction, now a displacement field, guarantees $\langle \ahat_l\rangle \neq 0$. Thus, a static `electric' field, proportional to $\gamma$, is ever present. Through hopping, this would generate an `Ising' interaction between the atomic dipoles. 
Moreover, a renormalized $\varepsilon$ would act as a `transverse' field on the local parities.   
Hence, the quantum Ising dynamics in the Rabi lattice model. Although the photon fluctuations above the static field would also be present, we can neglect them in the strong $\gamma$ limit.

To give this discussion a proper form, we transform $\Hhat_1$ under the displacement operator, $\Dhat=\prod_l e^{-\frac{\gamma}{\omega}(\ahat^\dag_l-\ahat^{ }_l)}$. Since $\Dhat^\dag \ahat_l \Dhat = \ahat_l-\frac{\gamma}{\omega}$ and $\Dhat^\dag \chihat_l\Dhat = \chihat_l \, e^{-\frac{2\gamma}{\omega}(\ahat^\dag_l-\ahat^{ }_l)}$, the `displaced' $\Hhat_1$ looks as follows. 
\begin{eqnarray}
&& \Dhat^\dag\Hhat_1\Dhat = \Hhat_2 = \sum_l\left[ \omega\left(\nhat_l+\frac{1}{2}\right) -\frac{\gamma^2}{\omega} \right] + \nonumber\\
&& \frac{\varepsilon}{2}e^{-\frac{2\gamma^2}{\omega^2}}\sum_l \chihat_l \, e^{-\frac{2\gamma}{\omega} \ahat^\dag_l}e^{\frac{2\gamma}{\omega} \ahat_l}\,\sigma^z_l -2t\frac{\gamma^2}{\omega^2}\sum_{l,\delta}\sigma^x_l\sigma^x_{l+\delta} \nonumber \\
&& -t\sum_{l,\delta}\sigma^x_l\sigma^x_{l+\delta} \left\{\left[\ahat^\dag_l\ahat_{l+\delta} -\frac{\gamma}{\omega}\left(\ahat^\dag_l+\ahat_{l+\delta}\right)\right] + h.c. \right\} \label{eq:H2}
\end{eqnarray} 
This is the RLM explicitly in terms of the static field, $\gamma/\omega$, and the photon fluctuations. 

Now invoke the strong coupling limit, $\gamma/\omega\gg 1$. That is, neglect the photon fluctuations in $\Hhat_2$. The limiting strong coupling problem, \( \Hhat^*_2 = -\frac{\gamma^2}{\omega}L + \omega\sum_l(\nhat_l+\frac{1}{2})+\frac{\varepsilon}{2}e^{-\frac{2\gamma^2}{\omega^2}}\sum_l\chihat_l\,\calL[{\nhat_l},\frac{4\gamma^2}{\omega^2}]\,\sigma^z_l -2t\frac{\gamma^2}{\omega^2}\sum_{l,\delta}\sigma^x_l\sigma^x_{l+\delta}\), is diagonal in photons, and has the structure of QI model, wherein for different photon occupancies $|\{n_l\}\rangle $, the dipoles and parities follow QI dynamics with differing transverse fields (and reference energies). This result is valid on a lattice of any form and dimension. Here, $\calL[n_l,x]$ is the Laguerre polynomial of order $n_l$ with $x=4\gamma^2/\omega^2$. For small enough hopping ($t\gamma^2/\omega^2 \ll \omega$), the lowest bunch of eigenstates of $\Hhat_2$ correspond to the $|\{n_l=0\}\rangle$ form of $\Hhat^*_2$, that is, the following QI model.
\begin{equation}
\Hhat_{QI} = e_0L + \frac{\varepsilon}{2}e^{-\frac{2\gamma^2}{\omega^2}}\sum_l\sigma^z_l -2t\frac{\gamma^2}{\omega^2}\sum_{l,\delta}\sigma^x_l\sigma^x_{l+\delta} \label{eq:HQI}
\end{equation}
Here, the Ising interaction, $J=-2t\frac{\gamma^2}{\omega^2}$, is ferro-electric (for $t>0$), the transverse field, $h=\frac{\varepsilon}{2} \exp{(-\frac{2\gamma^2}{\omega^2})}$, has an exponential reduction factor (akin to the Ham factors in vibronic states~\cite{Ham}), and $e_0=\frac{\omega^2-2\gamma^2}{2\omega}$ is a constant. Notably, when $t$ is only moderately small ($t\gamma^2/\omega^2\lesssim \omega$), the ground state of the RLM is still  described reasonably by $\Hhat_{QI}$. We have checked this, and the above proposition, about $\Hhat_{QI}$ by doing exact numerics on small clusters. For large $L$, the DMRG calculations presented below clearly confirm the strong coupling QI dynamics.

{\em Ground state properties.} 
The QI model is a well-studied problem. For the RLM ground state, it predicts two distinct phases. For $|J| \gtrsim h$, the Rabi lattice would exhibit spontaneous polarization. In this ferro-electric (FE) phase, the order parameter, $p = \frac{1}{L}\sum_l\langle\sigma^x_l\rangle \neq 0$. For $|J| \lesssim h$, each Rabi cavity would roughly behave as independent, and $p=0$ [para-electric (PE) phase]. The photon order parameter, $\psi=\langle\ahat_l\rangle$, after having applied $\Uhat$ and $\Dhat$, becomes $\psi=\langle \sigma^x_l(\ahat_l-\frac{\gamma}{\omega})\rangle \approx -p\frac{\gamma}{\omega}$. Thus, in the FE phase, the photons exhibit `superfluidity'. The PE phase may likewise be called a Mott phase. However, it not quite so because a Mott insulator is characterized by an integer polariton number per cavity, which is not true for the Rabi cavity. Moreover, the FE phase is not the usual superfluid described by a U(1) order parameter with gapless modes. Instead, it has an Ising (Z$_2$) order parameter with gapped excitations. Notably, the excitations in both the phases of QI model are gapped, except at the critical point where the gap closes continuously. 

The reduction factor, $\exp{(-\frac{2\gamma^2}{\omega^2})}$, renders the value of $\varepsilon$ unimportant, as $h$ would invariably be small for $\frac{\gamma}{\omega}\gtrsim 1$. 
Thus, it is immaterial whether $\varepsilon$ is near resonance or not. It also implies that the critical hopping, $t_c$, for the PE-FE transition would vary on an exponentially small scale.

To explicitly validate the strong coupling behavior, and to see deviations from it, we investigate the 1D Rabi lattice using DMRG. Since the 1D $\Hhat_{QI}$ is exactly soluble, it allows for proper comparisons with the DMRG results.

\begin{figure}[t] 
\includegraphics[width=6cm]{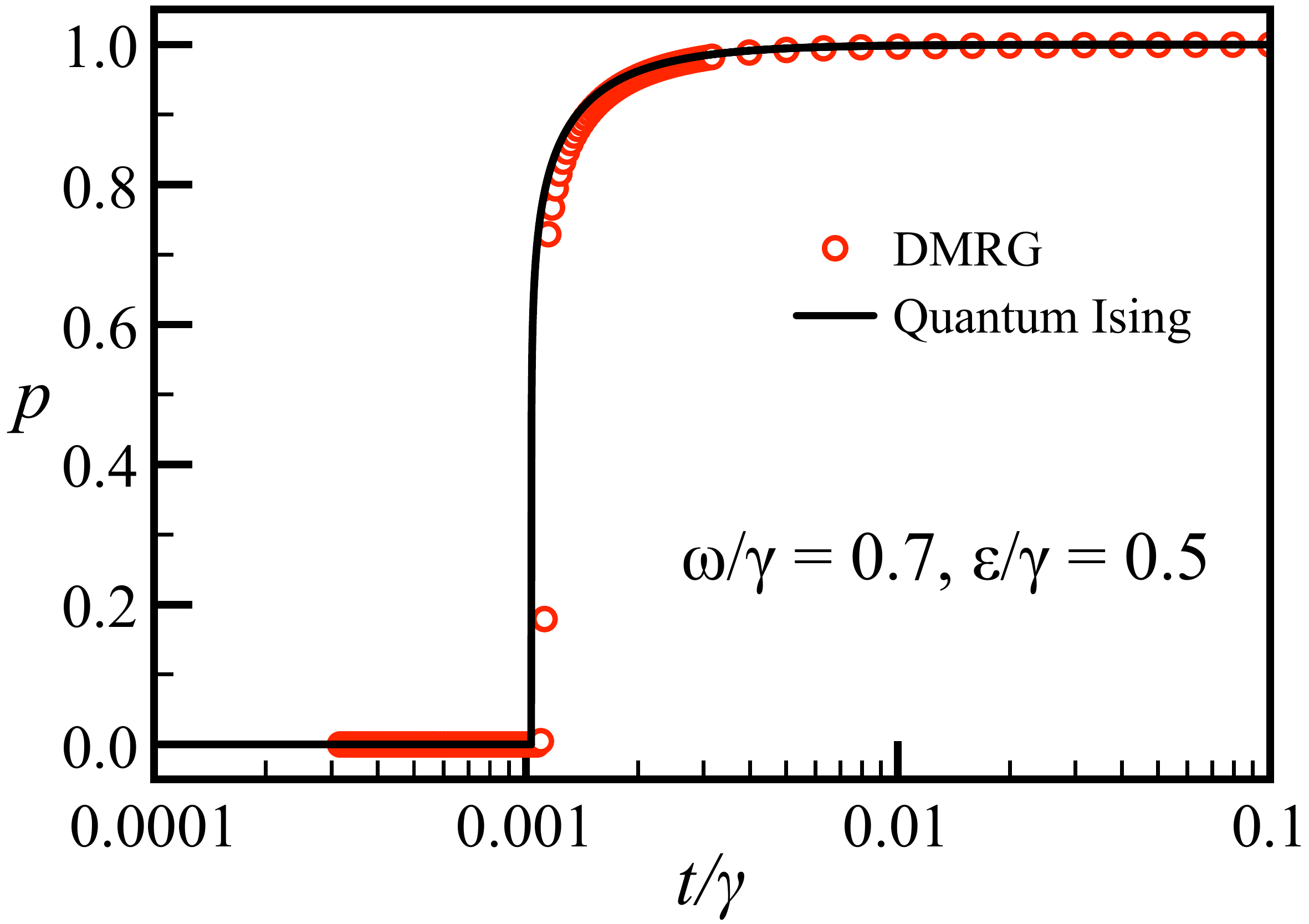} 
\caption{Polarization, $p$, vs. hopping, $t$, in the 1D Rabi lattice. The black line is the exact strong coupling prediction.}
\label{fig:mx_vs_t}
\end{figure}

For the 1D QI model, the exact polarization in the FE phase is $p = (1-\frac{1}{\lambda^2})^{\frac{1}{8}}$ (in the thermodynamic limit), where $\lambda=|J|/h$~\cite{QIsing.Pfeuty}. For Eq.~(\ref{eq:HQI}), $\lambda$ $=$ $\frac{4t}{\varepsilon} \frac{\gamma^2}{\omega^2} \exp{(\frac{2\gamma^2}{\omega^2})}$. Using DMRG, we compute $p$ as a function of $t$ in the ground state of RLM  (in the $\Hhat_2$ form). It exhibits PE to FE QPT, and agrees nicely with the strong coupling QI form (see Fig.~\ref{fig:mx_vs_t}). These calculations are done with four photon states ($n_l=0,1,2,3$) per cavity for $L$ upto 600.

\begin{figure}[htbp]
\includegraphics[width=6cm]{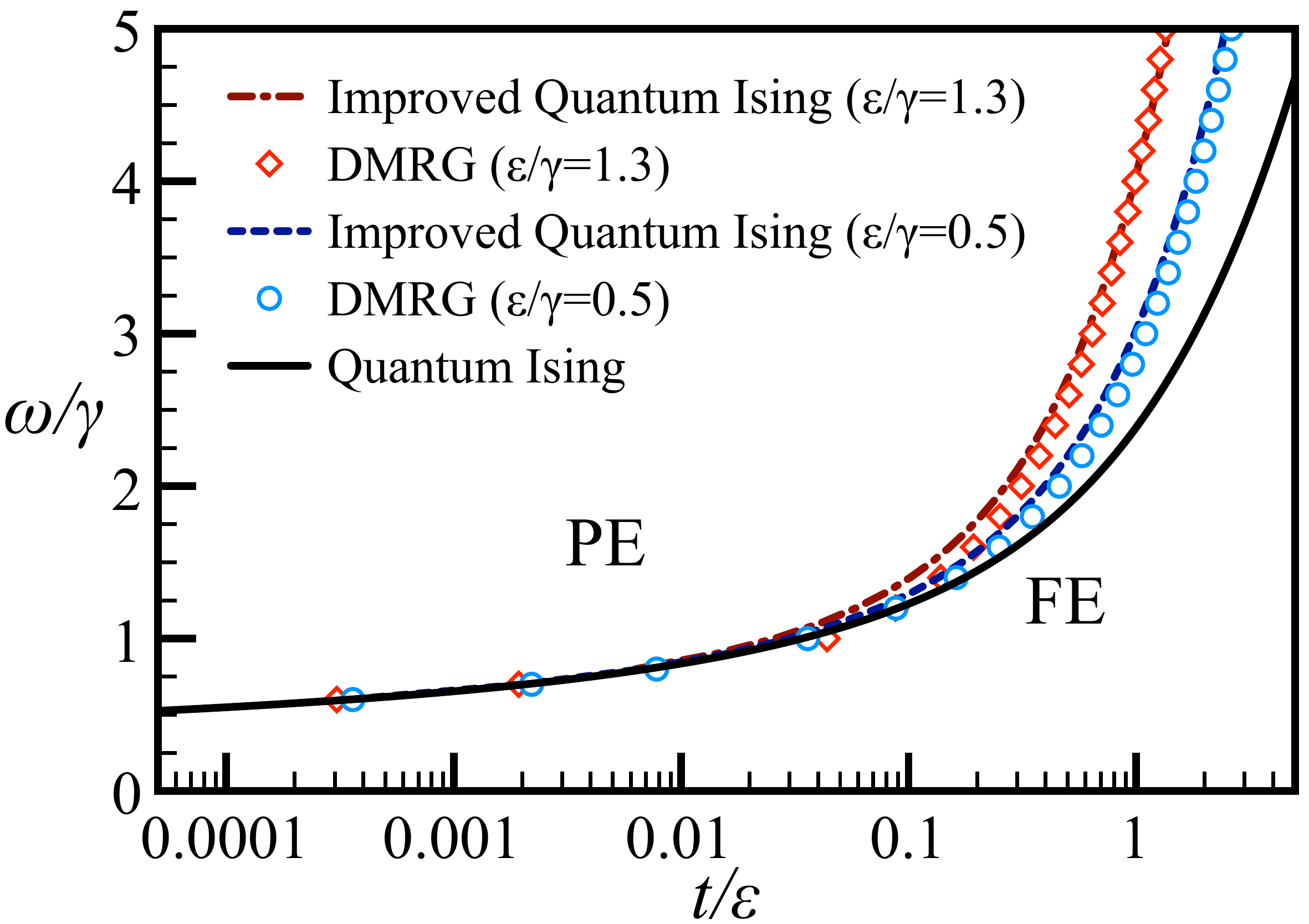} 
\caption{Quantum phase diagram for the 1D Rabi lattice. The PE (FE) denotes the para- (ferro-) electric phase.}
\label{fig:QPD}
\end{figure}

Since the exact quantum critical point for PE to FE transition in $\Hhat_{QI}$ in 1D is $\lambda=1$~\cite{QIsing.Pfeuty}, the strong coupling critical hopping is $t^*_c= \frac{\varepsilon}{4}\frac{\omega^2}{\gamma^2}\exp{(-\frac{2\gamma^2}{\omega^2})}$. For $\varepsilon=\omega$, the $t_c$ derived in Ref.~\onlinecite{RabiLattice.Schiro} is a particular case of our $t_c^*$. As the $t^*_c$  scales linearly with $\varepsilon$, in the plane of $\frac{t}{\varepsilon}$ and $\frac{\omega}{\gamma}$, the strong coupling phase boundary would be a universal curve, independent of $\varepsilon$ (in any dimension). In Fig.~\ref{fig:QPD}, the DMRG calculated phase boundaries are compared with the strong coupling QI prediction. For $\frac{\gamma}{\omega} \gtrsim 1$, the DMRG data indeed approaches the universal ${t^*_c}/{\varepsilon}$. Around $t_c$, it is also found that the gap $\sim |t-t_c|$, like the QI chain. 

For $\frac{\gamma}{\omega}\lesssim 1$, the deviations from the strong coupling QI behavior are approximately accounted for by an improved $\Hhat_{QI}$, wherein $\omega$ in Eq.~(\ref{eq:HQI}) is replaced by $\omega-zt\rho^x_1$. It corrects $\omega$ for photon dispersion. Here, $z$ is the nearest-neighbor coordination, and $\rho^x_1 = \langle \sigma^x_l\sigma^x_{l+\delta}\rangle$ is the nearest-neighbor dipole correlation. This improvement is effected by applying a displacement, $\prod_l e^{-\alpha(\ahat^\dag_l-\ahat_l)}$, on $\Hhat_1$, and  approximating $t\alpha\sum_{l,\delta}\sigma^x_l\sigma^x_{l+\delta} (\ahat_l+\ahat_{l+\delta}+h.c.)$ by $zt\alpha\rho^x_1\sum_l(\ahat^\dag_l+\ahat_l)$. Putting the linear photon terms to zero gives $\alpha=\gamma/(\omega-zt\rho^x_1)$, and ignoring the photon fluctuations gives the improved $\Hhat_{QI}$. The improved QI phase boundaries in Fig.~\ref{fig:QPD} are in good agreement with DMRG. For $\frac{\gamma}{\omega}\ll1$, it needs to be investigated further.

{\em Majorana-like edge modes.} 
On an open chain, the strong coupling $\Hhat_{QI}$ guarantees that the Rabi lattice has two edge modes in the FE phase. While the basic degrees of freedom in QI model (or RLM) are not fermions, the exact quasiparticles in the 1D QI model are fermions, of which, one consists of two Majorana-like  modes localized at the edges of the chain in the ordered phase~\cite{QIsing.Pfeuty,Kitaev.FreeMajorana}. Here, we address two points about these edge modes: 1) an observable signature, and 2) their stability against a `longitudinal' perturbation, $\sum_l\eta^{ }_l\sigma^x_l$. A stray electric field affecting the dipoles, $\sum_l E_l\sigma^x_l$, and the noise in the cavity mode, $\sum_l \mu_l(\ahat^\dag_l+\ahat^{ }_l)$, (after applying $\Uhat$ and $\Dhat$, and the strong $\gamma$ limit), are such perturbations. 

\begin{figure}[t] 
\includegraphics[width=6cm]{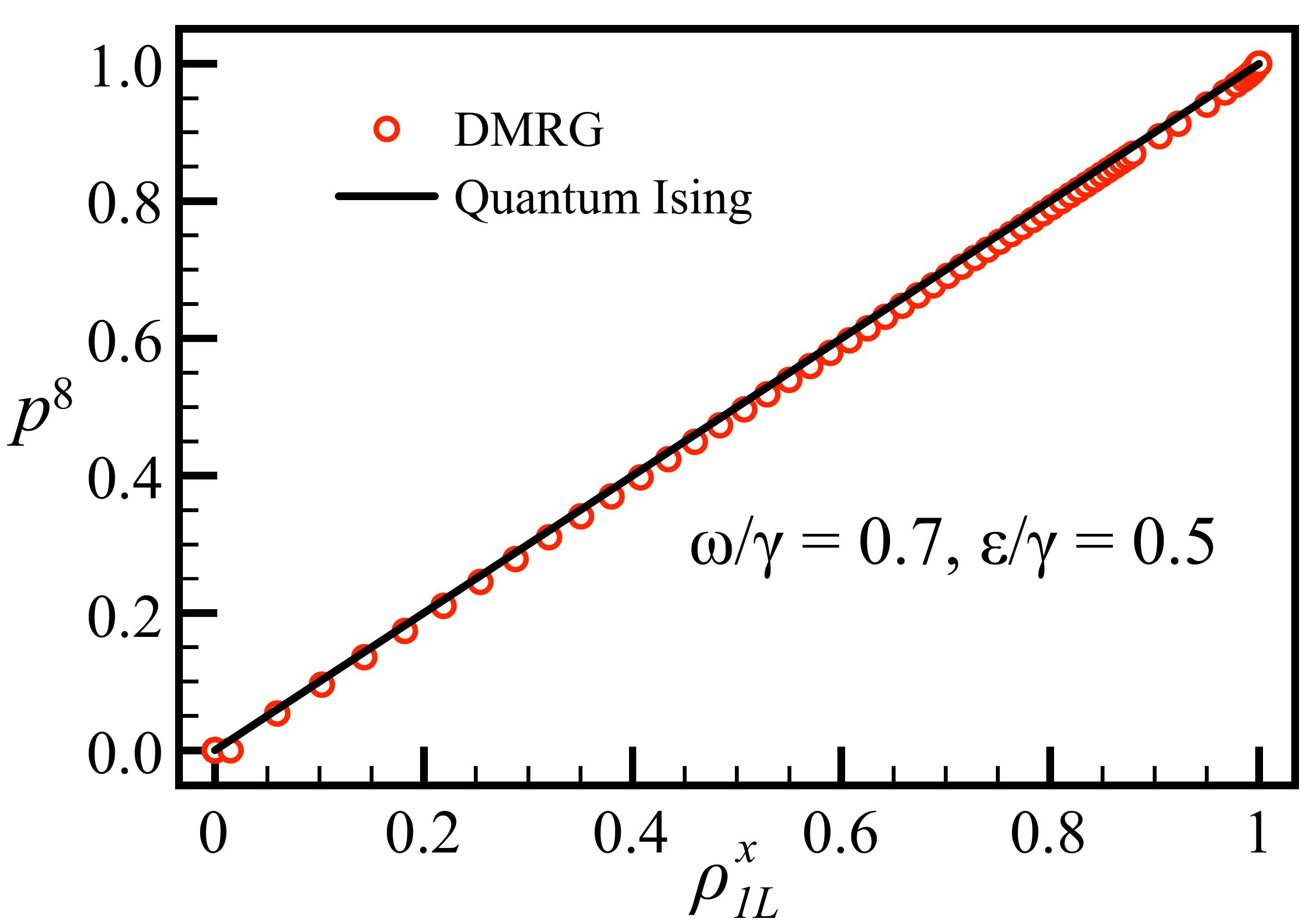} 
\caption{End-to-end correlation, $\rho^x_{1L}$, vs. $p^8$ in the ground state of the Rabi lattice model for $L=600$.}
\label{fig:end2end}
\end{figure}

A careful reading of Ref.~\onlinecite{QIsing.Pfeuty} reveals a special exact relation between the end-to-end correlation, $\rho^x_{1L}=\langle\sigma^x_1\sigma^x_L\rangle$, and the order parameter, $p$, in the ground state of an open QI chain. It is:  $\rho^x_{1L} = p^8+\mathcal{O}(1/L)$, different from the $p^2$ behavior of the long-range correlation in the bulk~\cite{Note.p2}. We realize this to be a consequence of the edge modes~\cite{LSM}. Hence, we propose $\rho^x_{1L}=p^8$ as a signature thereof. It relates two observables without explicitly involving model parameters. We do find this behavior in the 1D Rabi lattice. See Fig.~\ref{fig:end2end} (produced by varying $t$). The photon correlation, $\langle (\ahat^\dag_1+\ahat^{ }_1)(\ahat^\dag_L+\ahat^{ }_L)\rangle \approx \frac{4\gamma^2}{\omega^2} \rho^x_{1L}$, would also show $p^8$ behavior for a strong but fixed $\frac{\gamma}{\omega}$.
 
The twofold ground state degeneracy of the QI model for $\lambda>1$, in the thermodynamic limit, is due to the zero energy of the edge modes. For a finite $L$, this degeneracy is lifted by an amount $\lambda^{-L}$, causing tunneling between the edge modes~\cite{Topo_QC_Kitaev, QIsing.Pfeuty}. Since it is very small compared to the bulk quasiparticle gap for any finite $L\gg\frac{1}{\ln{\lambda}}$, the two modes effectively stay localized within a length of order $\frac{1}{\ln{\lambda}}$ at the edges, retaining their Majorana character. In view of this, we believe that $ \sum_l \eta_l \sigma^x_l$ will not be detrimental to the edge modes if $\eta_l = \frac{v_l}{L}$ (or weaker) for a $v_l \sim\mathcal{O}(1)$ and $\lesssim$ the gap. Deep inside the ordered phase ($\lambda\gg 1$), where the edge modes are highly localized, we expect them to better survive against non-zero $v_l$'s. Thus, $\eta_l \sim \frac{1}{L}$ is a `safe' perturbation on a finite $L$ array. Practically, a uniform $\eta=\frac{v}{L}$ can be viewed as an electric field  due to a potential drop $v$ along the array. 
 
\begin{figure}[t] 
\includegraphics[width=6cm]{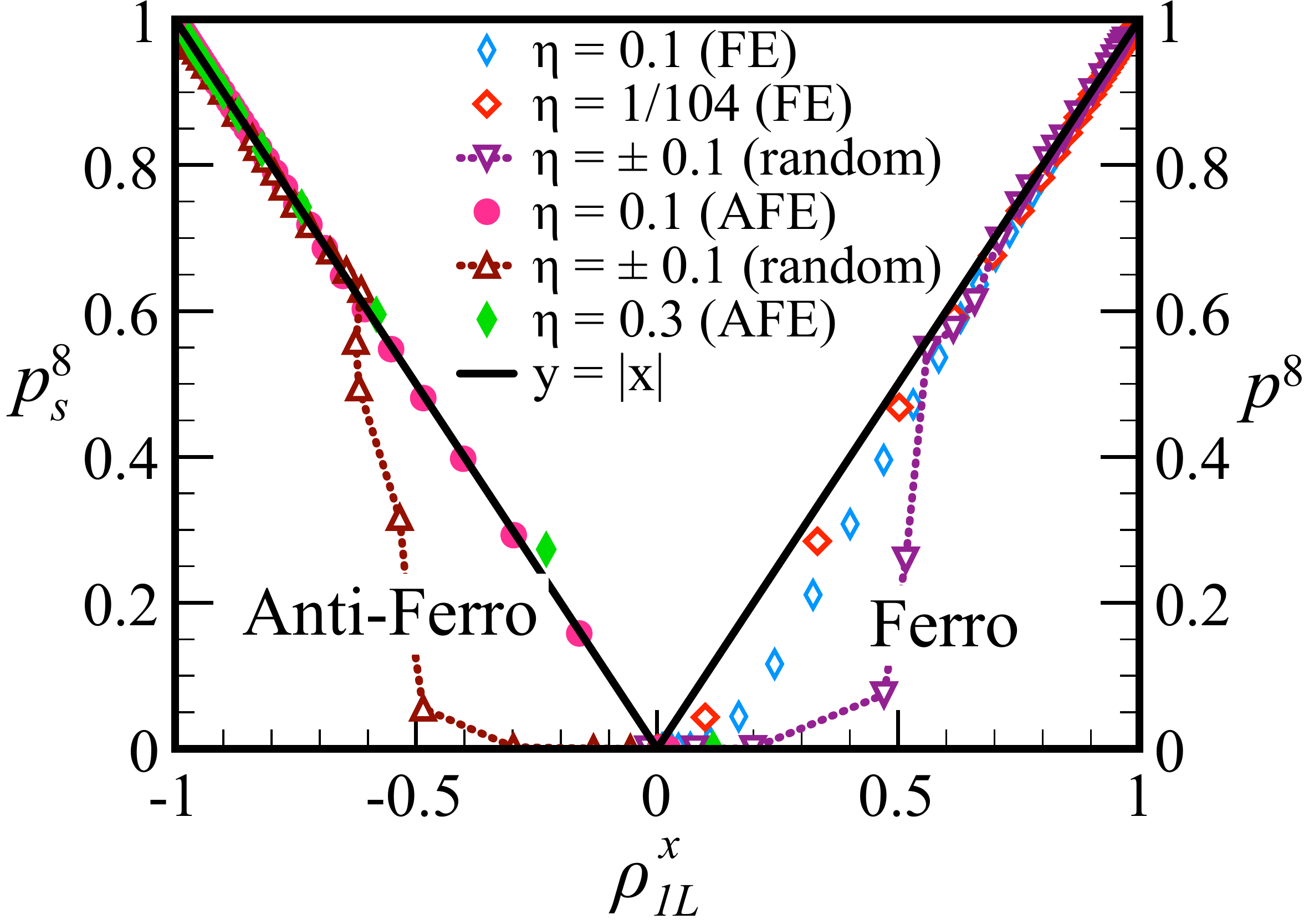} 
\caption{End-to-end correlation in the QI chain with $\eta\sum_l\sigma^x_l$. Except for $\eta= \frac{1}{104}$ ($L=104$, FE), all others have $L=400$. In the random case, $\eta_l$ is randomly $\pm 0.1$ (on every site), and the data is averaged over 100 ensembles. Here, $p_s$ is the staggered order parameter (for the anti-ferro $J$).}
\label{fig:end2endeta}
\end{figure}

To protect the edge modes from an $\eta$ stronger than $\frac{1}{L}$, having an anti-ferro $J$ (negative $t$) helps. As the ground state now is anti-FE ordered (for the nearest-neighbor case), its degeneracy is immune to a uniform $\eta\sim \mathcal{O}(1)$ sufficiently smaller than the gap. Moreover, a staggered $\eta^{ }_l\sim (-1)^l$ which can lift the degeneracy of the anti-FE ground state is not trivial like having a uniform $\eta$, and is probabilistically very unlikely (if $\eta_l$'s are randomly $\pm$). 

In Fig.~\ref{fig:end2endeta}, the DMRG data on the QI chain (produced by varying $|J|$ for $h=1$) supports our observations. Notably for strong $|J|$, the different cases of $\eta$ (including the random $\eta_l=\pm0.1$) fall on the exact line. Moreover, the anti-ferro case shows better protection against uniform $\eta$. It also applies to the strong coupling 1D RLM, but the values of $\eta$ rescale by $\frac{\varepsilon}{2} \exp{(-\frac{2\gamma^2}{\omega^2})}$, as the $h$ is $1$ in this data. For instance, $\eta=0.1$ in Fig.~\ref{fig:end2endeta} corresponds to $\frac{\eta}{\gamma} = .42 \times 10^{-3}$ for the RLM with $\frac{\omega}{\gamma}=0.7$ and $\frac{\varepsilon}{\gamma}=0.5$. The safe values of $\eta^{ }_l$'s for the RLM basically lie on the scale of critical hopping. Thus, the Majorana-like modes arising due to the QI dynamics are observable, and not as fragile as one believed, despite no topological protection.


{\em Conclusion.}   
For the emergence of QI dynamics in the Rabi lattice model, it is crucial not to enforce RWA on the dipole interaction. The strong coupling QI dynamics in the 1D Rabi lattice with open boundaries gives two Majorana-like edge modes in the ordered phase. These modes can be observed via the relation, $\rho^x_{1L}=p^8$, between the end-to-end dipole correlation and the order parameter. Even without the topological protection, the QI edge modes are found to be safe against the longitudinal field within some practical bounds (the anti-ferro order helps). This stability check for the QI edge modes is not exhaustive but a useful beginning into the hitherto unaddressed questions. It is necessary, for example, to also study the effects of frustration and dissipation on these edge modes, and to find ways to use them. Some of these studies will be discussed elsewhere~\cite{J1J2QI}.


{\em Acknowledgments.} 
B.K. thanks G. Baskaran and S. Girvin for useful comments, and acknowledges the visits to ICTP under the Associate scheme, during one of which the early ideas on this work were conceived. S.J. acknowledges CSIR-SRF for financial support. We thank Bimla Danu for discussions on DMRG. The DST-FIST support for the computing facility at SPS is also acknowledged.

\bibliography{manuscript.bib}

\begin{thebibliography}{39}
\bibitem [{not()}]{note.Majorana}%
  \BibitemOpen
  \href@noop {} {}\bibinfo {note} {For a fermion, created (annihilated) by
  $\chat^\dag$ ($\chat$), one can define two Majorana fermions $\phihat =
  \chat+\chat^\dag$ and $\psihat=i(\chat-\chat^\dag)$ such that
  $\phihat^\dag=\phihat$ and $\phihat^2=1$ (and likewise for $\psihat$). The
  Majorana operators $\phihat$ and $\psihat$ anti-commute with each other, and
  with other fermions.}\BibitemShut {Stop}%
\bibitem [{Not()}]{Note.p2}%
  \BibitemOpen
  \href@noop {} {}\bibinfo {note} {By definition, $\lim_{r\rightarrow\infty}
  \langle\sigma^x_l\sigma^x_{l+r}\rangle = p^2$. Here, $l+r$ and $l$ are two
  far far away points in the bulk that are also far away from the boundaries,
  for a thermodynamically large $L$.}\BibitemShut {Stop}%
\end{thebibliography}
\end{document}